\documentclass[%
 reprint,
 amsmath,amssymb,
 aps,
]{revtex4-2}

\usepackage{graphicx}
\usepackage{dcolumn}
\usepackage{siunitx}
\usepackage{tabularx}
\usepackage{bm}
\usepackage{caption}
\usepackage{subcaption}
\usepackage{threeparttable}

\begin{document}

\preprint{APS/123-QED}

\title{Special Glass Structures for First Principles Studies \\of Bulk Metallic Glasses}

\author{Siya Zhu}
 \email{siya\_zhu@brown.edu}

\affiliation{%
School of Engineering, Brown University, Providence, RI, 02912, USA
}%

\author{Jan Schroers}
\affiliation{Department of Mechanical Engineering and Materials Science, Yale University, New Haven, CT, 06511, USA}

\author{Stefano Curtarolo}
\affiliation{Department of Mechanical Engineering and Materials Science and Center for
Autonomous Materials Design, Duke University, Durham, NC, 27708, USA}

\author{Hagen Eckert}
\affiliation{Department of Mechanical Engineering and Materials Science and Center for
Autonomous Materials Design, Duke University, Durham, NC, 27708, USA}

\author{Axel van de Walle}%
\affiliation{%
School of Engineering, Brown University, Providence, RI 02912, USA
}%

\date{\today}

\begin{abstract}
The atomic-level structure of bulk metallic glasses is a key determinant of their properties. An accurate representation of amorphous systems in computational studies has traditionally required large supercells that are unfortunately computationally demanding to handle using the most accurate ab initio calculations. To address this, we propose to specifically design small-cell structures that best reproduce the local geometric descriptors (e.g., pairwise distances or bond angle distributions) of a large-cell simulation. We rely on molecular dynamics (MD) driven by empirical potentials to generate the target descriptors, while we use reverse Monte Carlo (RMC) methods to optimize the small-cell structure. The latter can then be used to determine mechanical and electronic properties using more accurate electronic structure calculations.
The method is implemented in the Metallic Amorphous Structures Toolkit (MAST) software package.

\end{abstract}

\maketitle


\section{\label{sec:introduction}Introduction}

For decades, bulk metallic glasses (BMGs) have been investigated for their excellent strength and fracture toughness \cite{schuh2007mechanical,yavari2007mechanical,eckert2007mechanical,greer2007bulk,ashby2006metallic,shao2020effect,neilson2015weibull}. Traditional amorphous alloys require high quenching rates to avoid crystallization, which restricts the size of powders and films \cite{schroers2010processing}. In contrast, the critical cooling rates of BMGs are much lower, allowing the size to be greater than \SI{1}{\milli\metre} in the smallest dimension, and promising the application as structural materials. \cite{inoue2000stabilization,lin1995formation,xu2004unusual} \par

The atomic structures of BMGs are of prime importance for predicting and explaining their macroscopic properties \cite{sheng2006atomic,cheng2009atomic,luo2004icosahedral}. Experimental characterizations with X-ray diffraction (XRD), small-angle neutron scattering(SANS), and transmission electron microscopy (TEM) help us understand some structural properties \cite{cheng2011atomic}, yet they are insufficient to fully reconstruct the atomic packing of BMGs. Molecular dynamics (MD) is often used in the investigation of atomic structures of BMGs \cite{duan2005molecular}. Traditional embedded atom method (EAM) MD can efficiently simulate the cooling process over a time scale of \SI{100}{\nano\second} for systems of over 10,000 atoms \cite{cheng2009atomic}. While this is typically sufficient to reproduce the general atomic structure of BMG,  EAM-MD exhibits two main shortcomings: 
First, due to the limited accuracy of EAM, the finer details of the structure, such as specific bond lengths or bond angles, may not be accurately reproduced. As a result, the accuracy of some physical properties, particularly viscosity and plastic mechanical properties, as well as normal mode frequencies, or even material density, may suffer.
Second, the structures obtained from EAM-MD cannot, due to their sizes, be directly used for further, more accurate analysis using first-principles calculations. This, therefore, precludes the calculations of electronic properties as well as more accurate calculations of mechanical properties.

To address this, it has been proposed \cite{sheng2006atomic} to use first-principles methods for both the simulation of the quenching process and the calculation of the properties of interest.
While first-principles methods guarantee a higher accuracy in calculating the energies and forces, the system size is typically limited to $<200$ atoms and the time scale to picoseconds.
As a result, the large statistical fluctuation and periodic boundary conditions in such a small system size as well as the high quenching rate due to the short simulation time can significantly affect the structure of the simulated system.

In an effort to leverage the advantages of both EAM-MD and fully first-principles methods, we propose to combine these methods with the help of Reverse Monte Carlo (RMC).
The idea of RMC has often been used to infer the atomic structures of BMG and other disordered systems \cite{pusztai1993structure,lamparter1995reverse,mcgreevy2001reverse} from indirect experimental characterization techniques. 
For instance, X-ray diffraction, neutron diffraction, extended X-ray absorption fine structure (EXAFS), etc., can provide structure factors, radial distribution functions, or atomic coordinations.
RMC modeling, a variation of the standard Metropolis Monte Carlo method, can then be used to generate structural model (a cell or an ensemble of atoms) by minimizing an error function (also called loss function) quantifying the consistency of the generated structure with the experimental data, while subjecting to a set of constraints like concentrations or cell shape. Various programs implementing this algorithm, such as fullrmc \cite{aoun2016fullrmc}, RMCProfile \cite{tucker2000simultaneous}, and RMC++ \cite{evrard2005reverse} are widely used to study the atomic modeling of liquids, glasses, polymers and other amorphous materials. 

Here we use RMC in differently. Instead of using experimental data as input, we perform RMC based on structures from EAM-MD results in order to build representative atomic structures that are small enough for DFT calculations, while maintaining the statistical geometrical properties of the large structure from EAM-MD.

We have developed a C++ software package --- Metallic Amorphous Structures Toolkit (MAST), as a part of Alloy Theoretic Automated Toolkit (ATAT) \cite{van2002alloy,van2017software}. MAST includes RMC codes for generating random structures suitable for MD and MC run, as well as scripts to automate the entire process, including MD, MC and ab initio calculations.
It also contains scripts for calculating mechanical properties such as elastic constants.

We first generate a typical BMG atomic configuration (with over 10,000 atoms) using  EAM-MD. Then, we use the RMC algorithm to construct a small periodic system of 32 atoms consistent with the MD result in terms of statistical properties such as pair distribution functions and bond angles. We give the name Special Glass Structure (SGS) to such atomic structure, paraphrasing the concept of Special Quasirandom Structures (SQS) \cite{zunger1990special,van2013efficient,saal2013thermodynamic} for crystalline structures. Once an optimal small-cell representation has been determined, ab initio calculations for structural optimization and mechanical properties computations are performed.

In this work, we study, as an example, Zr-Cu and Zr-Cu-Al based BMGs over a range of compositions, since these systems represent the transition metal BMG former and are the basis of various important BMGs with attractive properties \cite{xu2004bulk,wang2004bulk}. Atomic structures and mechanical properties are calculated and compared with experimental data and other theoretical predictions.

\section{\label{sec:methods}Methods}

\subsection{\label{sec:ErrorFunc}Error Functions}
To measure the structural differences between a large target structure with over 10,000 atoms and an SGS with 32 atoms, we devise an error function containing various kinds of commonly used structural information. \par

This information includes the pair distribution functions (also known as the partial radial distribution functions, PDF), defined as
\begin{equation}
    g_{\alpha \beta} \left( r \right)  = \frac{1}{4\pi r^2 \rho c_\beta }\frac{\text{d} n_{\alpha \beta}\left( r\right)}{\text{d}r}
\end{equation}
where $\rho$ is the atomic number density, $n_{\alpha\beta}\left( r \right)$ is the average number of atoms type $\beta$ within a distance of $r$ from an atom of type $\alpha$, and $c_\beta$ is the concentration of atom type $\beta$ in the structure. To make it symmetric to $\alpha$ and $\beta$, it can also be written as
\begin{equation}
    g_{\alpha \beta} \left( r \right)  = \frac{N}{4\pi r^2 \rho N_\alpha N_\beta} \frac{\text{d}N_{\alpha \beta}\left( r\right)}{\text{d}r}= \frac{V}{4\pi r^2 N_\alpha N_\beta} \frac{\text{d}N_{\alpha \beta}\left( r\right)}{\text{d}r}
\end{equation}
where $V$ is the volume, $N$ is the total number of atoms, $\text{d} N_{\alpha\beta}\left( r \right)$ is the total number of $\alpha$ and $\beta$ atoms pairs within distance $r$ to $r+\text{d}r$, $N_\alpha$ and $N_\beta$ are the total numbers of atoms type $\alpha$ and $\beta$, respectively. A traditional way to compare the PDFs is to obtain the structure factors with a Fourier transformation on the PDFs, sum over all partial structure factors for the total structure factors, and perform a $\chi ^2$ test with the one from experiments like XRD. Since we only compare between both atomic structures instead of experimental data, we can compute the difference between two PDFs directly. In order to get a smooth PDF without sharp peaks for small structures or noises for large systems, we perform kernel smoothing \cite{wand1995monographs}:
\begin{equation}
    \bar{g}_{\alpha \beta} \left( r \right) \equiv \frac{1}{s}\int_{-\infty}^{\infty} K\left(\frac{r'-r}{s}\right) g_{\alpha \beta} \left( r' \right) \text{d} r'
\end{equation}
where $K$ is a symmetric kernel function integrating to one and $s$ is a user-specified smoothing parameter.

Then, the error function for the PDFs is defined as
\begin{equation}
    E_{\text{PDF}} = \sum_{\alpha,\beta}\int_{0}^{r_\text{MAX}} \frac{1}{r^2}\left[ \bar{g}^\text{o}_{\alpha \beta} \left( r \right ) - \bar{g}_{\alpha \beta} \left( r \right)\right]^2 \text{d} r
\end{equation}
where $r_\text{MAX}$ is a cutoff distance (usually set as the distance of the 2nd or 3rd nearest neighbors), $\bar{g}^\text{o}_{\alpha \beta} \left( r \right )$ is the PDF of a target structure. We include a factor $1/r^2$ to give higher weights to shorter distances, as the chemical environment closer to an atom is more important than farther away. Notice that, since the PDF is symmetric in $\alpha$ and $\beta$, we do not double count the $g_{\beta\alpha}$ term in the sum. For example, for a Zr-Cu system, we only sum over three terms of partial PDFs: Zr-Zr, Zr-Cu, and Cu-Cu, but not Cu-Zr. For a ternary system like Zr-Cu-Al, we sum over 6 terms in $E_{\text{PDF}}$.
\par

For bond angles, we define a statistical quantity for the angles between two bonds connecting an atom $\beta$ to two nearest neighbors site occupied by species $\alpha$ and $\gamma$:
\begin{equation}
    h_{\alpha \beta \gamma}\left(\theta\right) = \frac{V N }{ N_\alpha N_\beta N_\gamma } \frac{\text{d}N_{\alpha \beta \gamma}\left( \theta \right)}{\text{d}\theta}
\end{equation}
where $\text{d}N_{\alpha \beta \gamma}\left( \theta \right)$ is the number of angles forming with bond $\alpha$-$\beta$ and $\beta$-$\gamma$, within the range of $\theta$ to $\theta+\text{d}\theta$. The bond length should be less than a cutoff length, usually set between the distance of the 1st and 2nd nearest neighbours. We also kernel smooth the function $h_{\alpha \beta \gamma}\left(\theta\right)$. The error function for angles is defined as
\begin{equation}
    E_{\text{angles}} = \sum_{\alpha, \beta, \gamma} \int_{0}^{\pi} \left[ \bar{h}^\text{o}_{\alpha \beta \gamma} \left( \theta \right ) - \bar{h}_{\alpha \beta \gamma} \left( \theta \right)\right]^2 \text{d} \theta
\end{equation}
Notice that $\alpha$ and $\gamma$ are equivalent and $\beta$ is not, therefore, there are 6 terms in the sum for a binary system and 18 terms for a ternary system. \par
For a total error function, we weigh the errors from each aspect:
\begin{equation}
    E = w_{\text{PDF}}E_{\text{PDF}} + w_{\text{angles}}E_{\text{angles}} 
    \label{equation:totalerror}
\end{equation}
A basic principle for choosing $w_i$ is to make every term $w_iE_i$ of the same order of magnitude, so that they are all taken into account in the Monte Carlo simulations.

\subsection{\label{sec:RMC}RMC process}
With an initial guess, which can be either a MD simulated structure, a previous MC result, or a random structure, we do an RMC process:
\begin{itemize}
    \item [\textbf{i.}]
    Randomly pick an operation from: (1) move an atom by a random amount; (2) swap positions of two atoms; (3) reshape the supercell by a random amount. The probabilities of each operation are chosen for better convergence. A suggested improvement is to set the initial structure as a cubic cell with volume based on the density of target structure, so that we do not need to reshape the cell (which is a default option in our code). Reshaping cells is sometimes detrimental to the convergence of the method and is not recommended;
    \item [\textbf{ii.}] 
    Calculate the error function $E_\text{new}$ for the new structure with the target structure; 
    \item [\textbf{iii.}] 
    Compare with the error function $E$ of the old structure. If $E_\text{new}<E$, we always accept the operation. Otherwise, we accept the operation with a probability $\text{exp}\left[ \left(E-E_\text{new}\right)/{T} \right]$, where $T$ is a ``temperature" factor greater than 0. The larger $T$ is, the more likely to get over barriers and find the global minimum, while a smaller $T$ speeds up the convergence to a local minimum. 
    \item [\textbf{iv.}] 
    Go back to the first step, until we reach the $E$ we set, or the maximum number of Monte Carlo steps. 
\end{itemize}

\subsection{\label{sec:generationmodel}Generation of atomic modeling}

As a specific example, to generate a 32-atom SGS with concentration Zr$_{50}$Cu$_{50}$, we follow this process:
\begin{itemize}
    \item [\textbf{i.}]
    Use MAST to generate a random structure with 8000 Zr atoms and 8000 Cu atoms;
    \item [\textbf{ii.}] Use LAMMPS \cite{plimpton1995fast,thompson2022lammps} to run an EAM-MD simulation. The EAM potentials in this work are from the group of Sheng \cite{cheng2009atomic}. We first do a NVT and a NPT run at \SI{2000}{\kelvin}, and quenching to \SI{300}{\kelvin} at $10^{12}$ K/s with NPT ensemble. Then a NPT at \SI{300}{\kelvin} is performed for a stable structure;
    \item [\textbf{iii.}] Use MAST for an RMC process from a cell with 16 Zr atoms and 16 Cu atoms randomly arranged to generate an SGS;
    \item [\textbf{iv.}] Use LAMMPS to do a structural optimization for the SGS from step(iii) with command ``minimize", then a NPT MD at \SI{300}{\kelvin}. This is because RMC may sometimes give unreasonable structures (with two atoms too close to each other). An optimization and MD run with EAM help us avoid this situation;
    \item [\textbf{v.}] Another RMC run on the result of step (iv). More loops of MD and RMC (step (iv) and step (v))can be performed for a better convergence. In this work, we do not perform extra loops of MD and RMC since it already gives very good fitting.
    \item [\textbf{vi.}] Use Vienna ab initio Simulation Package (VASP) \cite{kresse1993ab,kresse1996efficiency,kresse1996efficient} to do an ab initio optimization and get the final structure for Zr$_{16}$Cu$_{16}$. We use the Perdew-Burke-Ernzerhof (PBE) exchange and correlation functional at level of the generalized gradient approximation(GGA) \cite{perdew1996generalized} . A cutoff energy of 600eV is set to eliminate the effect of Pulay stress in structural optimization with ISIF=3. A $7\times7\times7$ K-points mesh is set. 
\end{itemize}

\subsection{\label{sec:elastic}Calculation of Elastic Moduli}
As an isotropic material, a BMG only has two independent elastic constants: $C_{11}$ and $C_{12}$. To calculate them, we apply strains on the relaxed structure, optimize with VASP (ISIF = 2), calculate the energy and fit the strain-energy relationship. A planewave energy cutoff of 300eV is used in the mechanical properties calculations and other settings are the same as previously stated. We apply uniaxial strain from $-0.03$ to $+0.03$ on $x$, $y$ and $z$ directions (the strain-energy curves on each direction are almost the same, which will be shown in the next section, as an evidence of isotropy of our structure). $C_{11}$ can be derived from
\begin{equation}
    C_{11} = \frac{2}{V_0}\frac{\text{d}^2E_\text{uniaxial}}{\text{d}\epsilon^2}
\end{equation}
Then we add a bulk strain $\epsilon$ from $-0.03$ to $+0.03$ on each direction and plot the strain-energy curve. Then $C_{12}$ can be calculated as
\begin{equation}
    C_{12} = \frac{1}{3V_0}\frac{\text{d}^2E_\text{triaxial}}{\text{d}\epsilon^2} - \frac{C_{11}}{2}
\end{equation}
The bulk modulus is 
\begin{equation}
    K = \frac{C_{11}}{3}+\frac{2C_{12}}{3}=\frac{2}{9V_0}\frac{\text{d}^2E_\text{triaxial}}{\text{d}\epsilon^2}
    \label{equation:bulkmodulus}
\end{equation}
and shear modulus is
\begin{equation}
    G = C_{44} = \frac{C_{11}-C_{12}}{2}
\end{equation}
\section{\label{sec:results}Results}
\begin{figure}[ht]
\includegraphics[width=\linewidth]{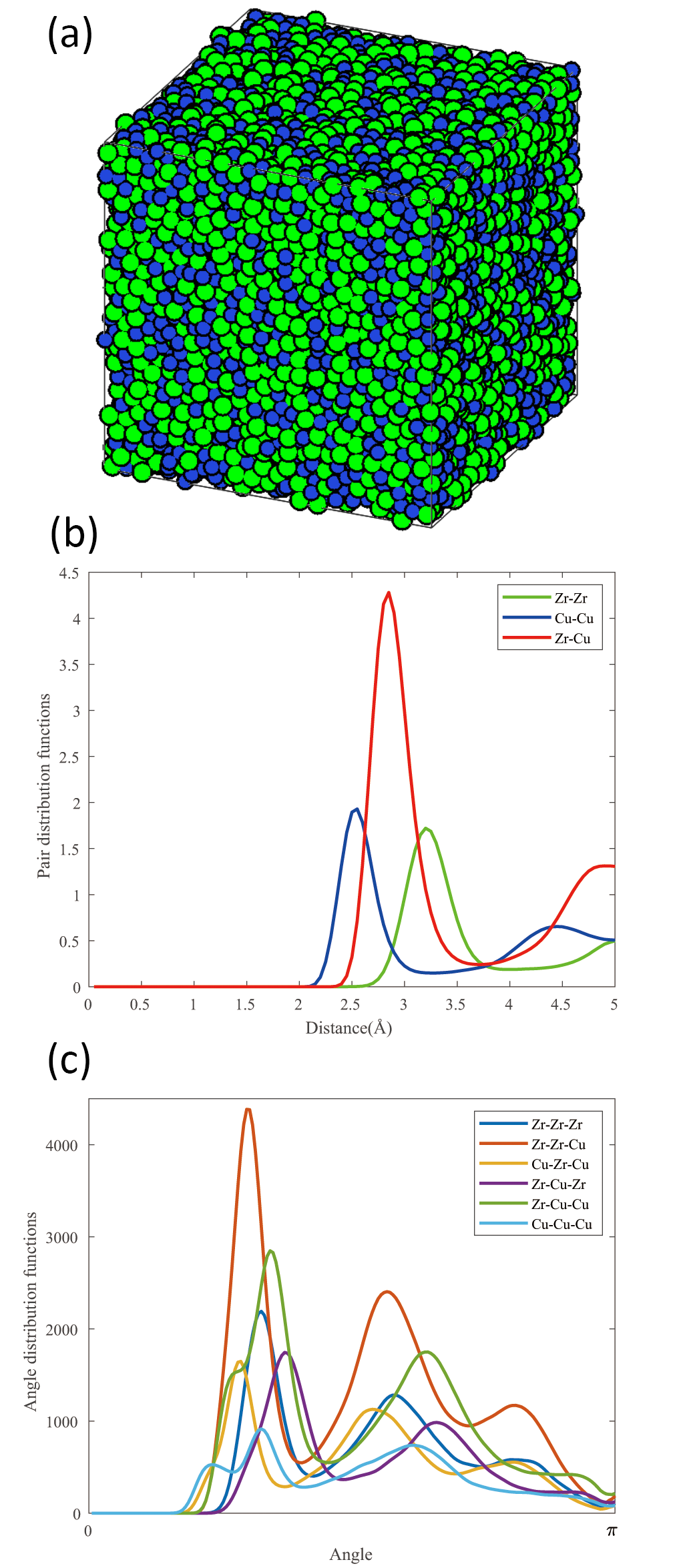}
			\caption{(a) An 16000-atom atomic structure of Zr$_{50}$Cu$_{50}$ constructed with MD. Green atoms are Zr and blue atoms are Cu. (b) Pair distribution functions of the structure.(c) Angle distribution functions of the structure, with cutoff distance 4A.} 
			\label{fig:fig1}
\end{figure}

In this section, we first use an example of Zr$_{50}$Cu$_{50}$ BMG to illustrate our results. First, we perform a MD simulation to generate a structure with 8000 Zr and 8000 Cu atoms, as shown in Figure \ref{fig:fig1}a. Pair distribution functions and angle distribution functions are calculated and shown in Figure \ref{fig:fig1}b and \ref{fig:fig1}c. \par
\begin{figure*}[b]
\includegraphics[width=\linewidth]{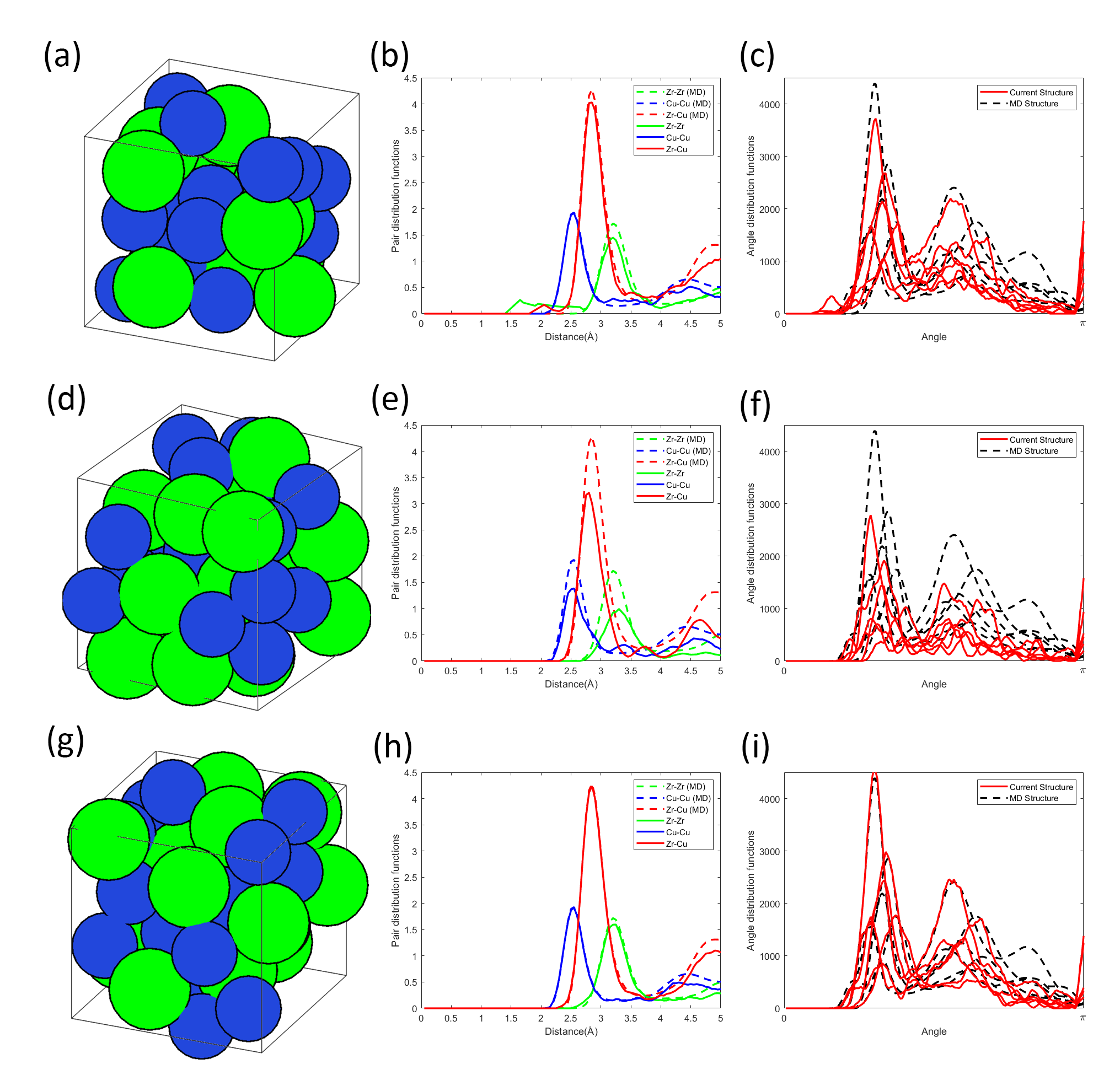}
			\caption{Atomic structures, PDFs and angle distributions at different steps in Section \ref{sec:generationmodel}. (a) SGS after step (iii); (b) PDFs for structure(a); (c) Angle distributions for structure(a); (d) SGS after step (iv);  (e) PDFs for structure(b); (f) Angle distributions for structure(b); (g) SGS after step (v); (h) PDFs for structure(c); (i) Angle distributions for structure(c).}
			\label{fig:fig2}
\end{figure*}

Following the steps in Section \ref{sec:generationmodel}, we generate the atomic structures with steps (iii) to (v), calculate pair and angle distribution functions as shown in Figure \ref{fig:fig2}.
The RMC steps sometimes generate structures with nonzero PDF at very short distances, thus implying that some atoms are nonphysically close to each other. This can be easily remedied either by performing relaxations and MD of the atoms using the EAM model, as illustrated in Figure \ref{fig:fig2}d-\ref{fig:fig2}f. Another RMC process reducing the error function gives an SGS Figure \ref{fig:fig2}g with good fitting, as shown in Figure \ref{fig:fig2}h and \ref{fig:fig2}i.  The SGS is now physically reasonable and successfully represents the structural information of the structure in Figure \ref{fig:fig1}a. 

\begin{figure}[h]
\includegraphics[width=\linewidth]{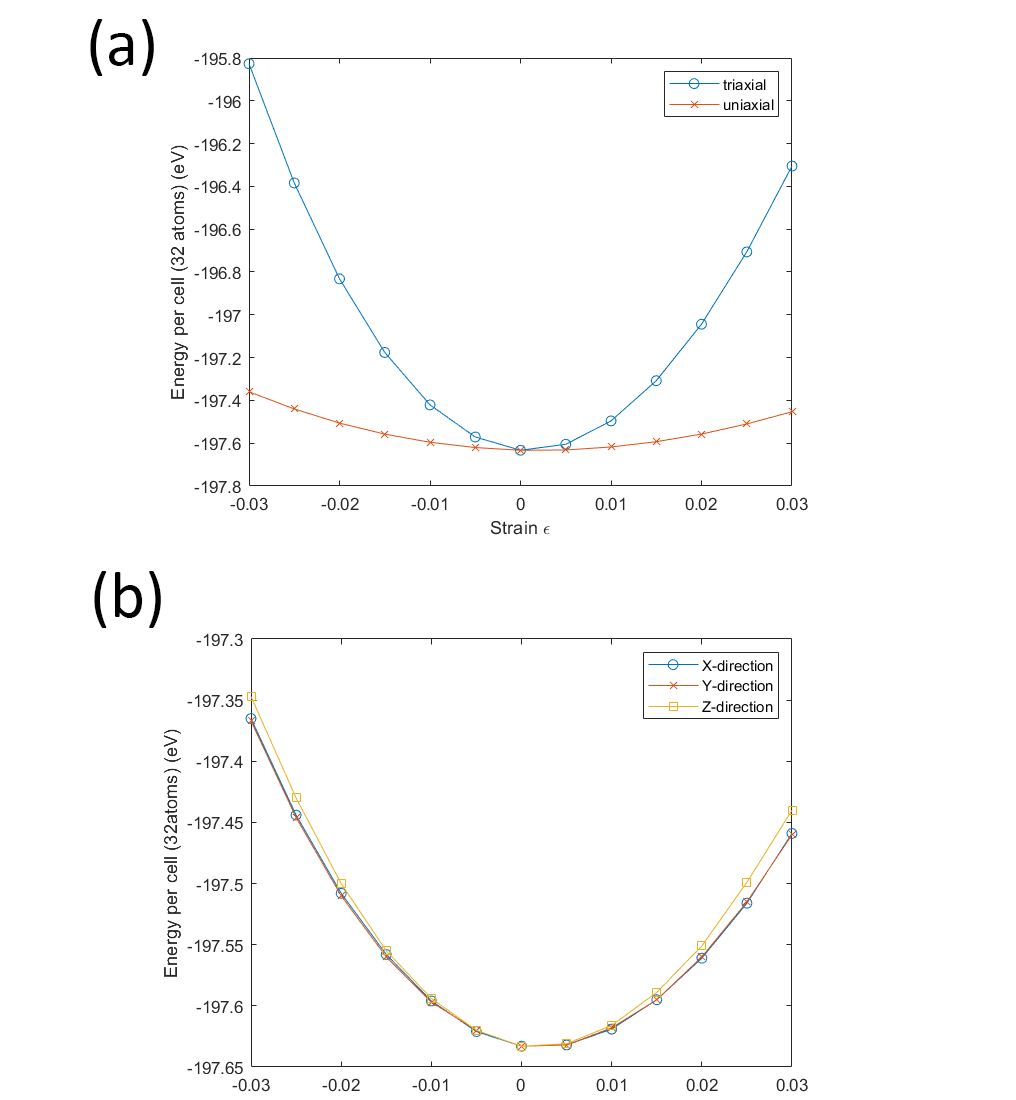}
			\caption{(a) Strain-energy curve of Zr$_{50}$Cu$_{50}$ SGS with uniaxial and triaxial strain; (b) Strain-energy curve of Zr$_{50}$Cu$_{50}$ SGS with uniaxial on $x$, $y$ and $z$ directions, respectively} 
			\label{fig:fig3}
\end{figure}

After the structural optimization with VASP, we apply uniaxial strain and triaxial strain to our SGS, optimize the structure under strain, and calculate the energy, as shown in Figure \ref{fig:fig3}. Two things need to be noted here: (i) we add the strain from $-0.03$ to $+0.03$, which might exceed the range of elastic deformation of BMG. However, in this case, the SGS is too small to perform any plastic deformation or fracture, and the energies still fit well with large strain. Therefore we keep the data of large strain up to 0.03 in this example. The strain range can be set manually in MAST; (ii) all ab initio calculations of energy here are at the temperature of \SI{0}{\kelvin}. That would bring about some error in the elastic modulus compared to experimental data at room temperature. Nevertheless, the differences between mechanical properties at \SI{0}{\kelvin} and \SI{300}{\kelvin} are insignificant. Figure \ref{fig:fig3}b shows the energy under uniaxial strain applied to $x$, $y$ and $z$ directions, respectively. The strain-energy curves of three directions almost overlap, manifesting that our SGS is ``isotropic" in some sense, which provides additional evidence of its validity. \par

In Table \ref{tab:table1}, we perform a robustness test of our method on 32 atoms structure of Zr-Cu BMG at concentration $x=0.5$. With four different initial structures in step (iii) in Section \ref{sec:generationmodel}, we get four different SGSs and calculate their mechanical properties. The results show our Monte Carlo process converges and is robust and reliable. \par
\begin{table}[b]
\caption{\label{tab:table1}Robustness test of MAST on Zr$_{50}$Cu$_{50}$ (32 atoms)}
\begin{ruledtabular}
\begin{tabular}{c | c c c c c c }
 &  C11(GPa) & C12(GPa) & K(GPa) & G(GPa) & E (GPa) & $\nu$\\ 
\hline
1 & 141.34 & 89.77 & 106.96 & 25.78 & 71.60 & 0.388\\
2 & 142.11 & 89.23 & 106.86 & 26.44 & 73.27 & 0.386\\
3 & 142.30 & 92.63 & 109.19 & 24.83 & 69.25 & 0.394\\
4 & 142.13 & 92.88 & 109.30 & 24.63 & 68.72 & 0.395\\
\end{tabular}
\end{ruledtabular}
\end{table}
In Table \ref{tab:table2}, we compare the mechanical properties of Zr$_{50}$Cu$_{50}$ from experiments, our method, and some other methods to show the advantage of our method. For the SGS in Figure 2(c) with 32 atoms, the differences in bulk modulus, shear modulus, Young's modulus and Poisson's ratios are less than 15\% (labeled MAST (32 atoms) in Table \ref{tab:table2}). For comparison, we apply strain on the structure with 16,000 atoms in Figure \ref{fig:fig1}a, run MD with EAM potentials for energies, and calculate mechanical properties(labeled as EAM-MD (16,000 atoms) in Table \ref{tab:table2}). The results from our process MAST agree with experimental data a little bit better --- both methods overestimate the bulk modulus K, while MAST underestimates shear and Young's modulus and EAM-MD overestimates them, compared with experiments. To show the beneficial contribution of our Monte Carlo process in MAST, we also show computations for the glass structure with 32 atoms, generated with EAM-MD only. We calculate the mechanical properties with EAM-MD itself(labeled as EAM-MD (32 atoms) in Table \ref{tab:table2}) and ab initio (labeled as EAM-MD + ab initio (32 atoms) in Table \ref{tab:table2}). Both of them show poor agreement with experiments. An SGS with only 12 atoms (Zr$_6$Cu$_6$) is also generated with MAST(labeled as MAST(12 atoms) in Table \ref{tab:table2}). The mechanical properties of it are close to MAST 32 atoms SGS, and are even better than EAM-MD structures with 32 atoms, which shows the robustness and accuracy of our method.\par
\begin{figure}[h]
\includegraphics[width=\linewidth]{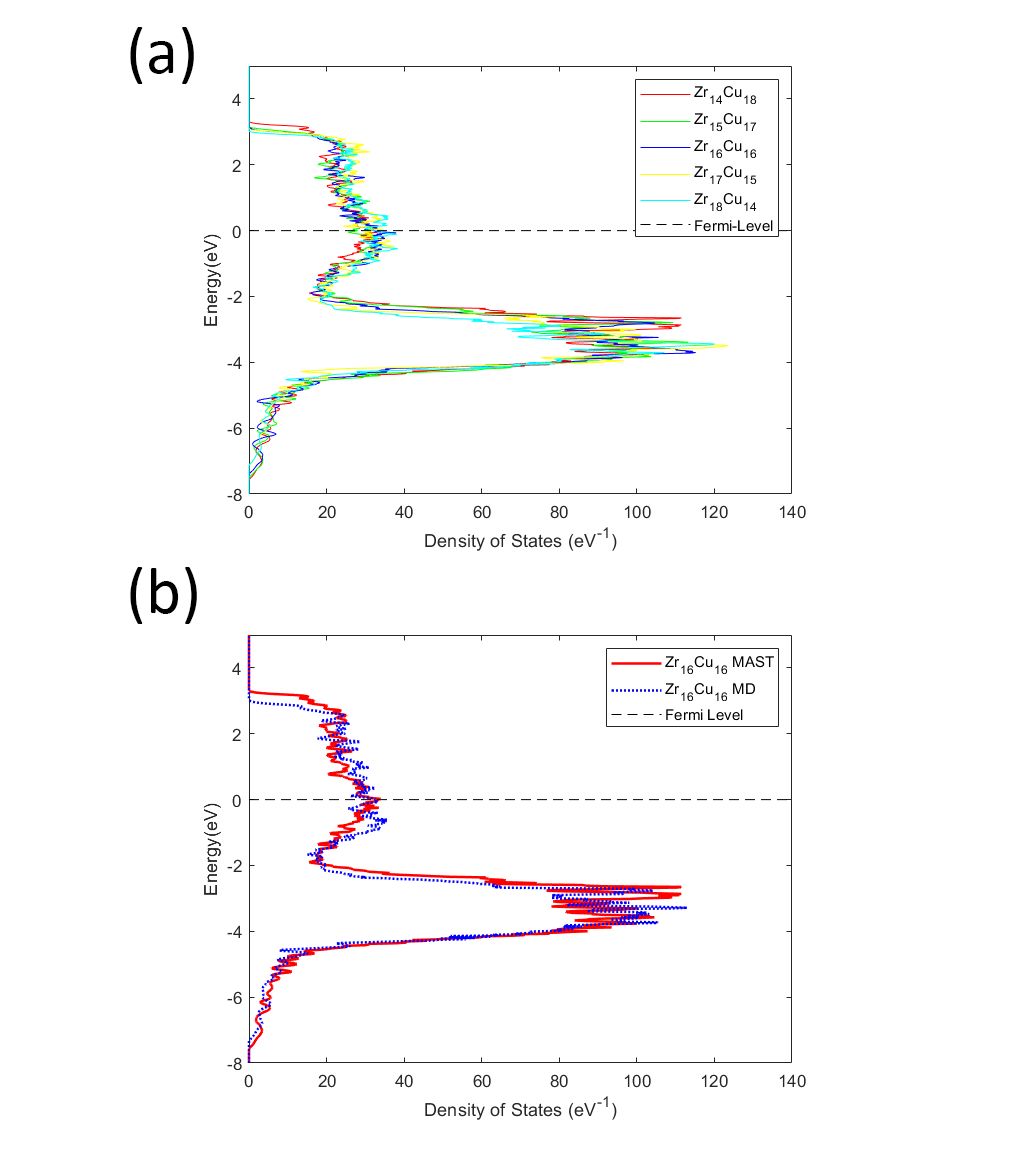}
    \caption{(a) DOS for ZrCu SGS generated by MAST at different concentrations; (b) DOS for Zr$_{16}$Cu$_{16}$ SGS by MAST and EAM.}
			\label{fig:fig4}
\end{figure}

\begin{table*}[b]
\caption{\label{tab:table2}Mechanical properties of Zr$_{50}$Cu$_{50}$ by different methods}
\begin{ruledtabular}
\begin{tabular}{c | c c c c c c}
&  K(GPa) & G(GPa) & E (GPa) & $\nu$ & $R^2$ & $R_{\text{adj}}^2$\\
\hline
Experiment \cite{fan2006thermophysical} & 101.2 & 31.3 & 84.0 & 0.35 \\
MAST(32 atoms, average) & 108.08 & 25.42 & 70.7 & 0.391 & $>$0.998 & $>$0.998 \\
MAST(12 atoms) & 111.9 & 24.5 & 68.6 & 0.398 & 0.987 &0.986 \\
EAM-MD (16,000 atoms) & 112.3 & 38.9 & 104.5 & 0.344 & 0.999 &0.998\\
EAM-MD + ab initio (32 atoms) & 107.3 & 15.9 & 45.3 & 0.430 & 0.991 & 0.990\\
EAM-MD (32 atoms) &110.6 & 40.7 & 108.7 & 0.336 & 0.926 &0.911\\
\end{tabular}
\end{ruledtabular}
\end{table*}
Similarly, our method can be applied to BMG systems with different concentrations, or even more than two elements. We show some calculation results for ZrCu and ZrCuAl BMG in Table \ref{tab:table3}, together with some experimental data as reference. For most of the concentrations, our results agree well with experiments on the bulk moduli, but the levels of agreement decrease for shear moduli and Young's moduli. This originates, in large part, from the fact that ab initio bulk modulus calculations exhibit a higher signal-to-noise ratio, because a triaxial strain induces a larger change in energy than a uniaxial strain and only the bulk modulus depends solely on triaxial strain data (c.f. Eq. \ref{equation:bulkmodulus}). As seen in Figure \ref{fig:fig3}, the overall energy changes associated with triaxial strains are about 10 times bigger than those associated with uniaxial strains.
In spite of this, the SGS approach still provides sufficient accuracy to investigate the compositional-dependence of elastic properties and, in this case, provides corroboration of the relatively small sensitivity to composition found in experiments.
\begin{table*}[b]
\caption{\label{tab:table3}Mechanical properties of ZrCu and ZrCuAl BMGs with different concentrations}
\begin{threeparttable}
\begin{ruledtabular}
\begin{tabular}{c | c c c c c c c c}
 & K(GPa,cal) & K(GPa,exp) & G(GPa,cal) & G(GPa,exp) & E (GPa,cal) & E(GPa,exp) & $\nu$(cal) &$\nu$(exp) \\
\hline
Zr$_{36}$Cu$_{64}$  \cite{xu2004bulk,johnson2005universal} & &104.3 & & 34 & & 92.3 & & 0.352\\
Zr$_{43.7}$Cu$_{56.3}$ & 112.9 & & 25.3 & & 70.6 & &  0.396 &\\
Zr$_{46.9}$Cu$_{53.1}$ & 112.8 & & 20.2 & & 57.1 & & 0.416 &\\
Zr$_{50}$Cu$_50$ \cite{fan2006thermophysical} & 108.8 & 101.2 & 25.4 & 31.3& 70.7 & 84.0& 0.391 &0.35 \\
Zr$_{53.1}$Cu$_{46.9}$ & 106.9  & & 18.5 & & 52.6 & & 0.418 & \\
Zr$_{56.3}$Cu$_{43.7}$ & 107.8  & & 21.1 & & 59.5 & & 0.408 & \\
Zr$_{54}$Cu$_{46}$ \cite{cheng2009correlation,johnson2005universal} & & 128.5 & & 30.0 & & 83.5 & & 0.391\\
Zr$_{47.5}$Cu$_{47.5}$Al$_5$ \cite{fan2006thermophysical}  & &  113.7 & & 33.0 & & 90.1 & & 0.365\\
Zr$_{47.5}$Cu$_{47.5}$Al$_5$ \cite{kumar2007plasticity} & & - & & - & & $\sim$60$^\dagger$& & -\\
Zr$_{47.5}$Cu$_{47.5}$Al$_5$ \cite{barekar2010structure} & & 117$\pm$1 & & 32$\pm$1 & & 88$\pm$1 & &-\\
Zr$_{47.5}$Cu$_{47.5}$Al$_5$ \cite{pauly2010criteria} & & - & & - & &87$\pm$5 & &-\\
Zr$_{47}$Cu$_{47}$Al$_6$ \cite{pauly2010criteria} & & - & & - & & 84$\pm$5 & &-\\
Zr$_{46.9}$Cu$_{46.9}$Al$_{6.2}$ & 108.2 & & 22.3 & & 62.6 & &  0.40 & \\
Zr$_{45}$Cu$_{45}$Al$_{10}$  \cite{thurnheer2014compositional} & & - & & - & & $\sim$60$^\dagger$ & &-\\
Zr$_{45}$Cu$_{45}$Al$_{10}$ \cite{kumar2007plasticity} & &- & &-& & $\sim$80$^\dagger$ & &-\\
\end{tabular}
\end{ruledtabular}
\begin{tablenotes}
\footnotesize
\item $^\dagger$ estimated from reported graph;
\end{tablenotes}
\end{threeparttable}
\end{table*}
The usefulness of SGS is not limited to mechanical properties.
For instance, they make it possible to analyze detailed electronic structure of BMGs, which would be impossible with EAM alone and intractable with large supercells using ab initio methods. In Figure \ref{fig:fig4}a we show the calculated DOS for different compositions of Zr-Cu glasses. This type of analysis enables us, for instance, to assess whether this BMG is well described by a rigid-band approximation. In the case of Zr-Cu, we can actually determine that shifts in the DOS as a function of composition, relative to the Fermi level, are not uniform over all energies. For instance, around 3eV, there is a larger shift between Zr$_{14}$Cu$_{18}$ and Zr$_{15}$Cu$_{17}$ composition than between Zr$_{15}$Cu$_{17}$ and Zr$_{18}$Cu$_{14}$ composition, while to opposite is true around $-2$ eV.\par
These effects would be difficult to isolate using other methods. For instance, in Figure \ref{fig:fig4}b, we make a comparison between the DOS obtained via an SGS from MAST and the one obtained via EAM-MD only, showing that the electronic properties would likely be incorrect without the Monte Carlo optimization process. The two DOS differ by shifts of up to 0.2 eV on the energy scale, which is large enough to mask the effects identified in Figure \ref{fig:fig4}a.
\section{Conclusion}
In this work, we develop a method to generate small-cell approximations to amorphous structures suitable for efficient ab initio calculations. These so-called Special Glass Structures (SGS) are constructed by matching geometric descriptors from large-cell empirical potential MD simulations using reverse Monte Carlo. The method is benchmarked by comparing the predicted mechanical properties of common Bulk Metallic Glasses to corresponding experimental results.\par
More generally, our method can be used for the study of any amorphous system, and is applicable to the calculation of properties besides the elastic constants, such as thermodynamics properties like enthalpy and vibrational entropy with a higher accuracy than classical potential-based MD. Larger SGS ($\sim 100$ atoms) should also apply to the study of kinetic processes, such as viscous flow or plastic deformation processes via shear band evolution in BMGs \cite{xie2008hardness}. While, we leave the demonstration of these capabilities for subsequent work, it should be clear that the SGS method opens the way to the systematic study of BMG properties via high-throughput ab initio methods over a broad range of chemistries \cite{van2017software,curtarolo:htnat}.

\medskip

\begin{acknowledgments}
This work is supported by Office of Naval Research grant N00014-20-1-2225. Computational resources were provided by (i) the Center for Computation and
Visualization at Brown University, (ii) the Extreme Science and
Engineering Discovery Environment (XSEDE)  through allocation TG-DMR050013N, which is supported by National Science Foundation Grant No. ACI-1548562 and (iii)
the Advanced Cyberinfrastructure Coordination Ecosystem: Services \& Support (ACCESS) program through allocation DMR010001, which is supported by National Science Foundation grants 2138259, 2138286, 2138307, 2137603, and 2138296.
\end{acknowledgments}

\end{document}